\documentclass[]{article}

\def\ben{\begin{equation}}
\def\een{\end{equation}}
\def\bea{\begin{eqnarray}}
\def\eea{\end{eqnarray}}
\input amssym.def
\input amssym.tex
\begin{document}

\hfuzz=100pt
\title{Causality and the Skyrme Model}
\author{G.W. Gibbons
\\
D.A.M.T.P.,
\\ Cambridge University,
\\ Wilberforce Road,
\\ Cambridge CB3 0WA,
 \\ U.K.}
\maketitle

\begin{abstract}
It is shown that despite the possibility of a breakdown
of hyperbolicity, the $SU(2)$ Skyrme model can never
exhibit bulk  violations of
Einstein  causality
because its energy momentum tensor satisfies the dominant
energy condition. It also satisfies the strong energy condition.
The Born-Infeld-Skyrme model also satisfies both the
dominant and strong energy
conditions.

\end{abstract}

%\tableofcontents
\vfill \eject

\section{Introduction}
The Skryme model continues to yield insights into pion and nucleon
physics \cite{Review}. Initially, the  static properties of the model were
studied but time-dependent processes are also important.
Their study
led to the realization that during the time evolution the
classical equations of motion, which are highly non-linear,
 may cease to be  hyperbolic
 \cite{CrutchfieldBell}. This is potentially worrying because
although the equations of motion are relativistically covariant, their
non-linear character makes
it unclear whether Einstein causality is
always maintained
during the time evolution. The danger is particularly great during
violent processes such as Skyrmion-anti-Skyrmion annihilation
\cite{Halasz}. Presumably if Einstein causality were to fail one would
have to give up the model, on the grounds that the various
approximations made to obtain it from microphysics, for example from
QCD, can no longer be valid, since those models incorporate Einstein
causality as a basic assumption. In this connection it  is
perhaps  interesting to
recall the suggestion that superluminal motion of pions
may be possible  in hadronic fluids \cite{BilicNikolic}
when one loop effects at finite temperature are taken into account
for the linear sigma model.

The point of this note is to show that, regardless of whether
hyperbolicity breaks down, the energy momentum tensor $T_{\mu \nu}$
of the $SU(2)$ Skyrme model satisfies the dominant energy condition
\cite{Hawking}  and
hence,
by a result of Hawking \cite{Hawking, HawkingEllis, Carter}, causal
behaviour
is guaranteed.

The dominant energy condition is a pointwise condition
on $T_{\mu \nu}$ and  states that $-T^\mu _\nu t^\nu $ is
future directed timelike or null  for every future directed timelike
vector $t^\mu$. Equivalently $T_{\mu \nu} t^\nu s^\mu \ge 0$, for all
pairs of future directed timelike vectors $t^\mu$ and $s^\mu$.
Another equivalent formulation is that in all  local Lorentz frames
$T_{00} \ge |T_{\mu \nu}|$ for all index pairs $\mu \nu$.
Yet another formulation is that if we can diagonalize $T_{\mu \nu}$
relative to the spacetime  metric $\eta_{\mu \nu}$, so that
$\eta _{\mu \nu} = {\rm diag } (-1,1,1,1)$ and $T_{\mu \nu}= {\rm
diag}
( T_{00}, T_{11}, T_{22}, T_{33})$, then
$ T_{00} \ge |T_{11}|$ etc. The quantities $T_{11}$ etc are called \lq
principle pressures \rq. Note that one cannot always diagonalize
an arbitrary symmetric tensor relative to $\eta_{\mu \nu}$
because it is indefinite, but it is possible in the generic case.

The intuitive idea behind this condition is
that if it holds, then relative to any local frame, the flow of energy
and momentum is future directed timelike or null. In other words,
locally matter flows no faster than light. Using this condition
Hawking was able to show, for example that if $T_{\mu \nu}$ vanishes
at one time outside  a compact set $K$, then $T_{\mu \nu}$ vanishes
 outside the future of $K$. In other words the local
condition guarantees that bulk motion is never superluminal.
In fact the main aim of Hawking's paper was rather broader.
He  showed  that classical field
theories satisfying the dominant energy condition do not permit
particle creation  {\it ex nihilo}
even if the background
metric is curved and time-dependent. In fact classical models of
particle creation in external fields
always seem to entail some loss of causality,
for instance particle and anti-particle appearing at spacelike
separation \cite{Zeldovich}. In the quantum theory of course, there is
no reason
to expect that the dominant energy condition continues to hold
and creation of particles can take place. In the case of the Skyrme
model one still expects topological conservation laws to hold however.
 In the context of quantum
gravity, this may  lead to some constraints on the number
of created
Skyrmions and whether it is odd or even \cite{Gib2}.

A remark that  will will prove useful in the  sequel   is
that if $T^1_{\mu \nu}$ and $T^2_{\mu \nu}$ are two energy momentum
tensors
satisfying the dominant energy condition, then so does $a T^1_{\mu \nu}
+  b T^2 _{\mu \nu}$, for positive $a$ and $b$. Put more formally, the
set
of energy momentum tensors satisfying the dominant energy condition
form a Lorentz-invariant  convex cone in
the space of all second rank symmetric tensors.

The Skyrme model is based on a map $\phi^A(t,{\bf x})$ from spacetime
to a target model $M$ with positive definite  metric $G_{AB}(\phi)$,
with $A=1,2,\dots, {\rm dim} \thinspace G$.
In the simplest case $M$ is $SU(2) \equiv S^3$
with the  bi-invariant or round metric. A more ambitious model takes
$M =SU(3)$.  The Lagrangian $L$ and the energy-momentum tensor $T_{\mu
\nu}$
are constructed from the pulled back metric $L_{\mu \nu}=G_{AB}(\phi)
\partial _\mu \phi ^A \partial _\nu \phi ^B$. Generically we can
diagonalize
$L_{\mu \nu}$ relative to $\eta_{\mu \nu}$. The eigenvalues are
necessarily
non-negative so we write them as $T_{\mu \nu}= {\rm diag} ( \lambda
_0^2, \lambda ^2_1,  \lambda _1^2, \lambda ^2_3 )$. The eigen-values
 $\lambda_1^2, \lambda ^2_2, \lambda ^2_3$ were introduced by Manton
in the static case for which $\lambda _0^2=0$ \cite{Manton}.

Consider, to begin with, a non-linear sigma model without Skyrme term.
This has
\ben
L^{\rm sigma} =-{ 1\over 2} L^\mu \thinspace _\mu, \qquad T^{\rm
sigma}
 _{\mu \nu}= L_{\mu \nu}
  +{1\over 2} L^\tau \thinspace _\tau  \eta _{\mu \nu}.
\een
In our special frame
\ben
T^{\rm sigma}
_{\mu \nu} = {\rm diag}{1 \over 2}
( \lambda _0^2 +\lambda_1 ^2 + \lambda
^2_2 + \lambda ^2_3,    \lambda _0^2 +\lambda_1 ^2 - \lambda
^2_2 - \lambda ^2_3,    \lambda _0^2 -\lambda_1 ^2 + \lambda
^2_2 - \lambda ^2_3,    \lambda _0^2 -\lambda_1 ^2 - \lambda
^2_2 + \lambda ^2_3). \label{5}
\een

By inspection,  the sigma model energy momentum tensor satisfies the
dominant energy condition. Although this guarantees causality, it certainly
does not guarantee non-singular evolution. Indeed it is known that
finite energy initial data can collapse to a  singularity in finite time
\cite{Shatah}. One even has an explicit solution illustrating this
collapse.
If $\chi, \alpha, \beta$ are polar coordinates on $S^3$, and $t,r,
\theta, \phi$ polar coordinates on spacetime, this solution
 is given, for $t<0$,
 by
$\chi = 2 \tan ^{-1} (-r/t)$  for $r \le -t$ and $\chi= {\pi \over 2} $ for
$r > -t$.
This represents a one parameter family of inverse
stereographic projections.

In order to prevent collapse, one adds an additional term, $L^{\rm Skyrme}$ to
the Lagrangian. For $M=SU(2)$, the Skyrme term may be written in
suitable units as
\ben
L^{\rm Skyrme} =L_ {\mu \nu} L^{\mu \nu} - L^\mu \thinspace
_\mu L ^\nu \thinspace _\nu.
\een
The contribution to the energy momentum tensor is
\ben
T^{\rm Skyrme} _{\mu \nu}=
 4 \Bigl ( L_{\mu \nu} L ^\sigma \thinspace L_\sigma -L_{\mu \lambda}
L_\nu \thinspace ^\lambda  \Bigr) + \eta _ {\mu \nu} L^{\rm Skyrme}.
\een
In the  adapted frame
\ben
T^{\rm Skryme}_{00}= 2\bigl(\lambda ^2_1 \lambda ^2_2 + \lambda _2^2
\lambda _3^2 + \lambda _3^2 \lambda _1^2 + \lambda _0^2 \lambda ^2 _1
+ \lambda _0 ^2 \lambda ^2_2 + \lambda ^2_0 \lambda ^2_3  \bigr ),
\label{1}\een
\ben
T^{\rm Skyrme}_{11}= 2\bigl (\lambda _1^2 \lambda ^2_2 +\lambda _1^2
\lambda ^2_3 - \lambda _2 ^2 \lambda ^2 _3 \bigr ) + 2\lambda _0^2
\bigl ( \lambda ^2_2 +\lambda ^2_3 -\lambda ^2_1 \bigr ).
\label{2}\een
Clearly  $T^{\rm Skyrme}_{00} \ge | T^{\rm Skyrme}_{11} |$,
which is sufficient to show that $T^{\rm Skyrme}_{\mu \nu}$ satisfies
the dominant energy condition, and hence by the convexity property, so
does
the complete energy momentum tensor $T_{\mu \nu}= T^{\rm sigma}_{\mu \nu}
+ T^{\rm Skyrme} _{\mu \nu}$.

Since,
\ben
L^\mu \thinspace _\mu =
-{T^{\rm sigma}}  ^ \mu _\mu, \qquad  L_{\mu \nu}=
{T^{\rm sigma}} _{\mu \nu}  -{1 \over 2} \eta _{\mu \nu}
{T^{\rm sigma}} ^\lambda _\lambda ,
\een
one has the nice formula
\ben
T^{\rm Skyrme}_{\mu \nu} =-4 T^{\rm sigma} _{\mu \lambda } {T^{\rm
sigma }} ^\lambda _\nu + \eta _{\mu \nu} T^{\rm sigma }_{\lambda \rho}
{T^{\rm sigma}}  ^{\lambda \rho}.
\een
However this does not immediately suggest a less basis dependent
proof that the dominant energy condition holds. An interesting
question
is whether the breakdown of hyperbolicity is associated with the
boundary of the cone. More practically, it might be illuminating
to calculate the eigen-values
$\lambda^2_0, \lambda^2 _1 ,\lambda^2  _2,
 \lambda^2 _3 $ during the numerical evolution
as a diagnostic tool.

From (\ref{1}, \ref{2}) one
\ben
T^{\rm Skyrme} _{00}+ T^{\rm Skyrme}_{11} + T^{\rm Skyrme}_{22} +
T^{\rm Skyrme}_{33} \ge 0. \label{3}
\een
One easily checks from (\ref{5})
that (\ref{3}) also holds for the sigma model
energy momentum tensor. It follows that $T_{\mu \nu}$ belongs to the
relativistically invariant convex cone of symmetric second rank
tensors for which
\ben
\bigl ( T_{\mu \nu} + {1 \over 2} \eta _{\mu \nu} T^\lambda _\lambda
\bigr ) t^\mu t^\nu \ge 0,
\een
for all future directed timelike or null vectors $t^\mu$.
In other words, the strong energy condition \cite{HawkingEllis}
is also  satisfied. Physically this means that the gravitational field of
Skyrmion matter is always  attractive.

A different extension of the non-linear sigma model Lagrangian
to include higher powers of first derivatives is the
Born-Infeld-Skyrme
model. This has
\ben
L^{\rm Born-Infeld} = \sqrt{-\det \eta _{\mu \nu}}
- \sqrt{-\det( \eta _{\mu \nu} + L_{\mu \nu} )}.
\een

One finds that

\ben
T^{\rm Born-Infeld} _{00}=  { 1 \over \sqrt {1-\lambda _0^2 }} \Bigl (
\sqrt{ (1+\lambda ^2_1)(1+ \lambda ^2_2) (1+\lambda ^2_3) }-1 \Bigr ),
\label{6}\een
and
\ben
T^{\rm Born-Infeld}_{11}= 1- {\sqrt{1-\lambda ^2_0}\over 1+\lambda
^2_1} \sqrt {(1+\lambda ^2_1)(1+\lambda ^2_2)(1+\lambda ^2_3)}.
\label{7}\een

Note that the arbitrary constant in the Lagrangian has been chosen to
make it and the energy momentum tensor vanish for a
constant
field configuration.
It follows  from (\ref{6},\ref{7}) that $T^{\rm Born-Infeld}_{00} \ge
|T^{\rm Born-Infeld}_{11}|$, and so the Born-Infeld-Skyrme energy
momentum tensor also satisfies the dominant energy condition.
However, unlike the usual Skyrme model, a scaling argument reveals
there are no  smooth
static solutions with positive energy.

Finally we note that it follows easily from (\ref{6}) and (\ref{7})
that  the Born-Infeld-Skryme energy momentum tensor
also  satisfies the strong energy condition. Thus neither Skyrme matter
nor Born-Infeld-Skyrme matter is a suitable candidate for the dark
energy currently expanding the universe.
 To violate the strong energy condition one may shift the arbitrary
 constant mutiple of $g_{\mu \nu}$
in $T_{\mu \nu}$. This looks less arbitrary if one also introduces a
potential
as one does in the case of tachyons and considers
\ben
L=-V(\phi) \sqrt {\det (g_{\mu \nu} +L_{\mu \nu} )}.
\een
Now one finds that this can lead  to a
period of
acceleration \cite{Gib}
but as it stands this is probably not a viable model either for the
inflaton or for the dark energy.

\section{Acknowledgements} Some of the ideas in this paper emerged
during unpublished work  with Michael Green. I would like to thank him
and
Nick Manton for many helpful and enjoyable discussions.

\end{document}